# Performance limit of on-chip speckle spectrometers

BHUPESH KUMAR[1], GRAHAM D. BRUCE[1], LUCA DAL NEGRO[2,3,4] AND SEBASTIAN A. SCHULZ[1,5,*]

[1]SUPA, School of Physics and Astronomy, University of St Andrews, North Haugh, St Andrews KY16 9SS, UK
[2]Department of Electrical & Computer Engineering and Photonics Center, Boston University, 8 Saint
Mary's Street, Boston, Massachusetts 02215, USA
[3]Division of Material Science and Engineering, Boston University, 15 Saint Mary's Street, Brookline, Massachusetts 02446, USA
[4]Department of Physics, Boston University, 590 Commonwealth Avenue, Boston, Massachusetts 02215, USA
[5]Department of Electrical and Computer Engineering, Waterloo Institute of Nanotechnology, University of Waterloo, 200 University Avenue W, Waterloo, ON N2L 3G1, Canada.
* sas35@st-andrews.ac.uk,
sebastian.schulz@uwaterloo.ca

**Abstract:** Disorder-driven, integrated speckle spectrometers offer exceptional spectral resolution within a compact design. They benefit from enhanced optical path lengths due to multiple light scattering events, however, often at the cost of low optical throughput. Here, we investigate the relationship between these two figures of merit by systematically varying the scattering strength of random-uniform disorder distributions. Furthermore, we also investigate the temperature stability of such spectrometers. Our study shows that the device resolution can be tuned from 2 nm to 20 pm, while the operating temperature ranges from 1 to more than 6 degrees and throughput can be varied by more than a factor of 10, paving the way for application-tailored design of microscale high-resolution spectrometers.

**Introduction:** The ability to precisely measure an optical spectrum is key to many applications, such as absorption spectroscopy, the characterisation of optical and photonic devices and fundamental science from astronomy to biophotonics, quantum optics and engineering. Conventional spectrometers and spectrum analysers separate light of different wavelengths into different propagation angles, via diffraction or refraction, creating a one-to-one mapping of detector position to wavelengths. Thus, their performance is limited through strict trade-offs. To increase the resolution, either the propagation distance must be increased, resulting in a larger spatial separation of closely spaced wavelengths, at the cost of increased footprint, or the diffraction strength must be increased, e.g. by narrowing a spectrometer slit, resulting in a higher angular spread, but at the cost of reduced operating bandwidth and signal strength.

Therefore, significant research effort is focused on miniaturizing spectrometer systems [1–3] while maintaining the potential for high-resolution and broadband operation, including integrated spectrometers based on arrayed waveguide (AWG) gratings [4–6], echelle gratings or superprisms [7–9], as well as metasurface-based spectrometers [1, 10–12]. In such spectrometers, high-index waveguides or AWGs can increase the relative phase accumulated per unit length, thus shrinking the device footprint. However, this comes at the cost of reduced bandwidth, while the fundamental principle remains the same: increasing the resolution of an AWG- or superprism-based spectrometer requires either a larger footprint or a reduced bandwidth [13]. Thus, a fundamentally different spectrometer design is required to overcome these trade-offs: speckle-based spectrometers.

In a speckle-based spectrometer, the interference of many optical paths is used to generate a wavelength-dependent optical speckle [14], which in turn can be used to reconstruct the input spectrum [15–17]. Such speckle spectrometers can be generated in bulk optical settings, e.g. using optical fibres or integrating spheres [18, 19], but also in an integrated on-chip fashion, again through interference in multi-modal spiral waveguides, but also multiple scattering of light in highly disordered systems [17, 20 –23]. This disorder-driven spectrometer approach has the potential for the most compact spectrometer format, with typical devices having a footprint on the order of 0.0025 mm$^2$ [24]. Furthermore, in these structures, any wavelength of light that is not absorbed will result in a speckle pattern at the output section of the device. Therefore, when broadband transparent materials such as silicon nitride are employed, a single spectrometer can be operated over more than 1000 nm spectral bandwidth [25, 26].

However, these devices still feature significant challenges and their associated performance trade-offs. In disorder-driven spectrometers, the speckle pattern is generated through the interference of different optical paths generated by multiple scattering of light. While this allows us to increase the optical path difference (the key physical determinant of the device resolution) by increasing the scattering strength and reducing the mean free path, every scattering event also leads to optical loss as some component of light is scattered out of the plane of propagation. Thus, we introduce a new trade-off, where a higher resolution is associated with stronger out-of-plane scattering, reducing the device throughput and, hence, the sensitivity/signal-to-noise ratio. To date, this trade-off has not been fully explored. Initial studies show that an appropriate choice of scatterer distribution, for example, golden angle spirals [17] or multi-fractal patterns [27], increase the device throughput without affecting the spectral coherence - commonly seen as a stand-in for resolution - of the resulting speckle patterns. Yet these studies also point to another challenge for disorder-driven spectrometers: unlike conventional structures, where the resolution can be predicted from a simple diffraction calculation/characterisation, in a speckle spectrometer the resolution is dependent on the device, experimental noise and the spectral reconstruction algorithm employed, and characterising this property requires extensive measurements and numerical calculations.

In this paper, we explore the impact of scattering strength on optical throughput, spectral resolution and operating temperature ranges for disorder-driven on-chip spectrometers with a random uniform disorder distribution.

Finally, we will then show how the temperature-induced variation of the speckle pattern manifests itself and its effect on the spectrometer operation.

Our spectrometer design follows the principles of refs [17, 27] and is shown in Fig. 1. It consists of a two-dimensional, semi-circular scattering region of uniformly random distribution of air holes in the top Si layer of a 220 nm silicon-on-insulator (SOI) wafer. In this region, light will undergo multiple scattering events, leading to the formation of the speckle pattern. The straight edge of this scattering region is bound by a photonic crystal (PhC) mirror to minimize light scattering into the backward direction. A defect waveguide within the PhC guides light into the scattering region. The light that reaches the outer, circular edge of the semicircular scattering region hits a semicircular air trench, which redirects some of it out of plane for capture by an infrared camera. This fabrication process and the methodology for generating the disordered patterns are described in detail in the Methods section. We control the scattering strength by adjusting the density of scattering particles within the disordered area, from 1% to 10%, and observed a significant improvement in spectral resolution, by two orders of magnitude from approximately 2 nm to 20 pm. By studying the associated reduction in device throughput with increasing scatterer density and temperature-dependent device performance, we clarify the relationship between scattering characteristics and spectrometer performance.

## 2.2. Numerical Simulation

We performed 3D finite-difference time-domain (FDTD) simulations for each device to analyse the in-plane and out-of-plane scattering. Due to the high computational requirements of FDTD simulations, we perform these for a scaled-down version of the device, with a reduced scattering region with a diameter of 20 $\mu$m, comprising a disordered array of air holes, each with a diameter of 150 nm. We simulated all 10 devices over the wavelength range of 1550 nm to 1575 nm, recording the fields in 0.2 nm wavelength intervals. Figure 2(a, b) illustrates the out-of-plane scattering and in-plane field distribution at a scatterer density of 1% for a wavelength of 1550 nm, while panels (c) and (d) display the corresponding results for a 10% scatterer density. From both the out-of-plane and in-plane scattering, we observe that the lower, 1 % disorder density retains a strong component of light that propagates in the forward direction, either through ballistic transport or single scattering events with only a small angular redirection. Additionally, a tree-branch pattern forms on either side of this forward axis, caused by light undergoing a few scattering events. The 10% disorder density, however, results in considerably more diffuse transport, with the vast majority of optical paths being formed through many scattering events, including evidence of backwards propagating paths. As expected, the optical throughput is dependent on disorder density, both as a function of the wavelength (Fig. 3(a)) and as the total

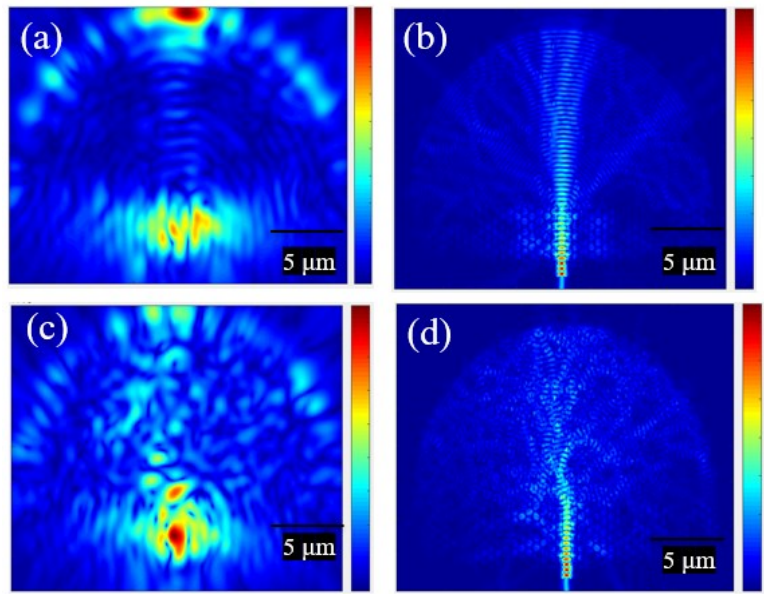

Fig. 2. (a, b) 3D Numerically simulated out-of-plane and in-plane scattering respectively from a disorder spectrometer device having 1% of scatterer density. (c, d) 3D Numerically simulated out-of-plane and in-plane scattering respectively from a disorder spectrometer device having 10 % of scatterer density.

device throughput, averaged across all wavelengths, shown in Fig. 3(b).

### 2.3. Experimental Calibration and Spectral Reconstruction

For the experimental characterisation, we obtain a reference transmission matrix by recording the speckle pattern at the scattering trench for each device over a representative wavelength range from 1550 nm to 1565 nm, using a tunable narrow-band (linewidth: 1 pm) laser. For the lower density devices (1 % -7 %), this scan is performed with a step size of 0.05 nm, which is reduced to 0.01 nm for the high density (8 %-10 %) devices, to ensure the proper characterisation of the device performance. We subsequently repeat this process to obtain multiple "measurement" matrices, which differ from the "reference" matrix through the variation in random noise inherent to any repeat optical characterisation, to assess the spectrometer's performance under realistic noise conditions.

Fig. 4(a, c) shows exemplary images of the device and speckle patterns recorded at a wavelength of 1550 nm for the devices with scatterer density 1 % and 10 %, respectively. For each wavelength step, the light scattered from the outer air trench is isolated, divided into 25 equal parts, and compiled as one column in the transmission matrix (T), see Fig. 4(b, d). Here, we can see

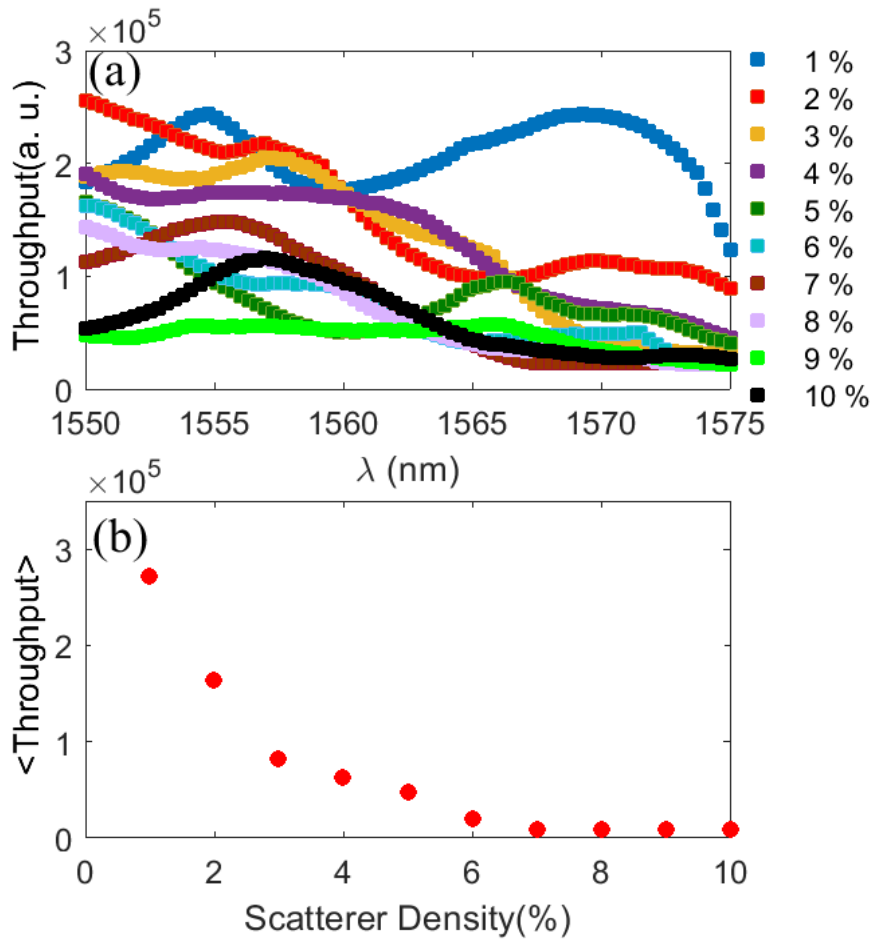

Fig. 3. (a) 3D FDTD simulated throughput (integrated light power scattered out-of-plane along the air) for all scatterer densities with, plotted as a function of wavelength in the range 1550 nm to 1575 nm. (b) Average throughput (averaged across all wavelengths) plotted as a function of scatterer density.

behaviour similar to the numerical simulation of the reduced area devices. The output of the

1 % scatterer density device is mostly concentrated close to the straight-through axis of the device (detectors 10-15 in panel (d)), indicating a strong ballistic transport component, while the output for the 10% scatterer density device (panel (b)) features a much wider spatial distribution. Similarly, a visual inspection of panels (b) and (d) shows a faster spectral variation of the speckle pattern for the higher scattering density, indicating that this should result in a higher resolution. Fig. 4(e) illustrates the throughput (the integral of the collected light over the detector region) for the different scatterer densities. As discussed earlier, an increased scatterer density should result in more scattering events and, thus, more out-of-plane loss and reduced throughput. Fig. 4(f) shows the average throughput over the scanned wavelength as a function of scatterer density. From both (e) and (f) it is evident that higher scatterer densities significantly reduce total throughput, from $2.9 \times 10^5$ counts for a 1% scatterer density to $3.6 \times 10^4$ counts for a 10% scattering density. To characterise the spectrometer resolution, we test the device under two conditions. First, we obtain an output spectrum when a single probe wavelength is incident, i.e. with the device acting as a wavemeter. Here, a narrower full-width half maximum of the spectral peak indicates better performance. We also test the device with multiple input wavelengths to see the closest wavelength spacing for which the different spectral components can be resolved. In all cases, the probe signals come from the "measurement" matrices, while the transmission matrix used for the retrieval algorithm (see "Methods" section) is the "reference" measurement. Therefore, this indicates the device's performance under realistic experimental conditions and accounting for experimental noise. Fig. 5 panels (a, c, e) show the wavemeter performance for devices with air holes density of 1 % (a, b), 5 % (c, d) and 10 % (e, f), respectively. Notably, as the air hole density increases, the half width at half maximum (HWHM) of the reconstructed wavelengths peaks decreases. This behaviour is also reproduced in panels (b, d, f), which show the spectrometer operation, specifically the closest spacing between wavelength peaks that could be resolved. Here, the increasing scattering density results in improved resolution, with the highest resolution of 20 pm, achieved for the 10 % scattering density device. We also investigated the thermal stability of the speckle spectrometers. For this, we use the same "reference" matrix as earlier, recorded at temperature $T_0$; however, now we measure additional "probe" matrices recorded with a known temperature offset $\delta T$, by placing the sample on a temperature-controlled stage. We then test the wavemeter performance of the spectrometer in the presence of this temperature difference between the reference and probe measurements. Our results show that the primary effect is a spectral shift of the retrieved wavelength, and in Fig. 6, we show both the average shift for different disorder densities and temperature offsets, as well as the standard deviation of this shift. We can observe that as the temperature offset increases, both the standard deviation and the wavelength offset increase. This behaviour seems independent of the scatterer density, with all devices featuring a shift of 0.31 - 0.33 nm for a 5°C temperature offset.

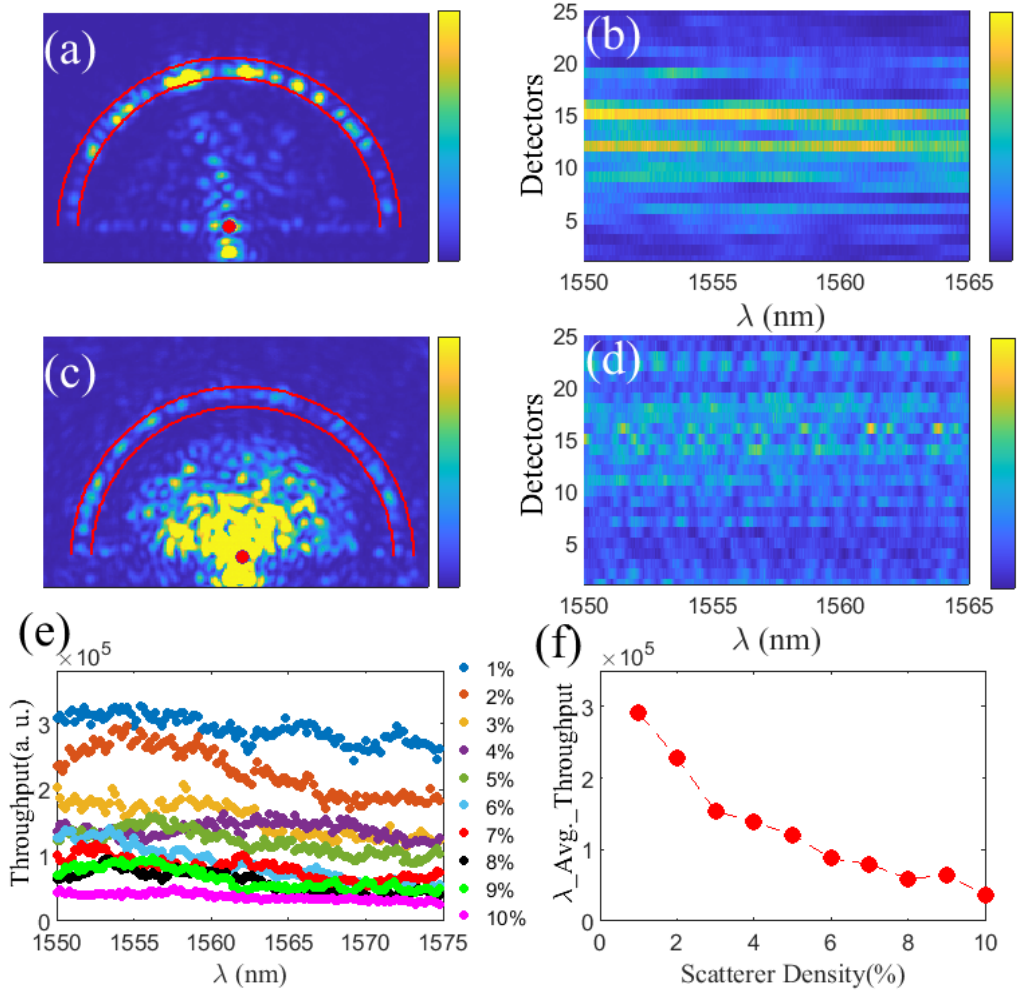

Fig. 4. (a), (c) are experimentally measured near field images of the random spectrometers with scattering density 1 % and 10 %, respectively. Images were recorded at a wavelength of 1500 nm. (b) and (d) shows the Calibration matrix measured over a wavelength range of 1550 nm to 1565 nm for devices with scatterer density 1 % and 10 % respectively Note wavelength step size in (b) is 0.05 nm, and in (d) it is 0.01 nm. (e) Throughout, i.e., the light scattered from the air trench, integrated along the air trench, as a function of wavelength. (f) Averaged (across all wavelengths) throughput plotted as a function of scatterer density.

## 3. Discussion

Our results confirm the initial statement that the throughput and resolution of these speckle spectrometers are highly dependent on the scattering density. Specifically, the resolution and resolving power increase as the scattering density increases, while the throughput decreases. This is clearly indicated in Fig. 7. Here, we show the resolving power instead of the resolution to visualize the trade-off between these two-performance metrics better, as both FOMs should be maximised. We note that while the throughput drops by roughly 1 order of magnitude (from $2.9 \times 10^5$ to $3.6 \times 10^4$ counts), the resolving power improves by two orders of magnitude, from

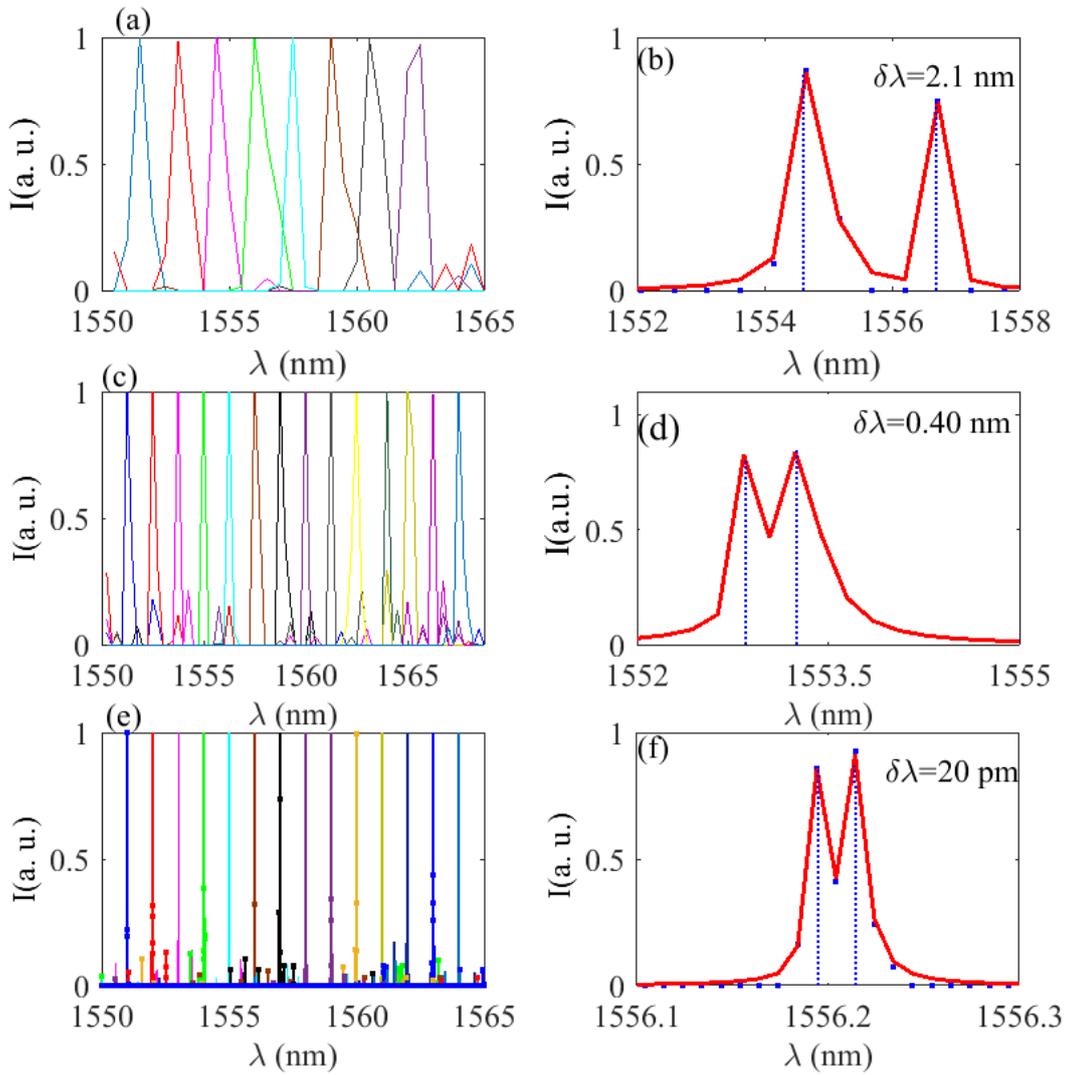

Fig. 5. (a, c, e) Reconstructed spectral lines for different arbitrary probe wavelengths using the TM matrix of the device having air holes density of 1%, 5%, and 10%, respectively. (b, d, f) The Spectral reconstruction of two simultaneous probe wavelengths for the devices with air holes density of 1%, 5% and 10%, respectively.

775 to 77500, corresponding to the two orders of magnitude improvement in the resolution from 2.1 nm to 20 pm. As mentioned earlier, when discussing Fig. 2, the low scattering density seems to be dominated by ballistic and singularly scattered light transport, which would indeed be expected to result in a low-resolution spectrometer, while the higher densities showed evidence of multiple scattering and multi-path interference.

From the thermal stability measurements presented earlier (Fig. 6) we can see the first-order approximation commonly made in integrated optics, i.e. that a temperature shift will result in a near linear shift of the operating wavelength of a device, also applies to disorder-driven spectrometers, with this shift dominated by the temperature variation, rather than the nature of the device. However, due to the complex nature of the speckle patterns in these spectrometers, this approximation is an oversimplification. With increasing temperature difference, the standard deviation of the shift observed for each measurement also increases, which can drastically affect the device's operation. Thus, if the wavelength shift is predictable (i.e. small error bar) it can be accounted for during the wavelength retrieval calculation as long as the temperature/wavelength shift is known. However, as the standard deviation of the wavelength shift increases and approaches or exceeds the HWHM of the spectrometer resolution, this can no longer be accounted for during the retrieval step. Therefore, we here define the usable temperature range for a disorder-driven spectrometer as that where the standard deviation of the temperature-induced wavelength shift, $\sigma_{\Delta\lambda T}$, is smaller than the spectrometer resolution, $\delta\lambda$. Since the resolution is strongly dependent on the scattering density, this also extends to the usable temperature range.

In Fig. 9 we therefore show the ratio between the temperature-induced wavelength shift and the spectrometer resolution ($\frac{\sigma_{\Delta\lambda T}}{\delta\lambda}$) for devices with a scatterer density of 1 %, 5 %, and 10 %. We can see that this ratio always remains far below unity for the lowest scatterer density (1 %), i.e., the variation in wavelength shift is always smaller than the resolution of the spectrometer. As such, this spectrometer can be used over the full temperature range investigated here and most likely for a significantly larger range, as the ratio is below 0.2 even for a 6°C temperature variation. For the device with 5 % scattering density, the ratio approaches unity for the 6°C temperature variation, indicating the maximal temperature range over which this can be used without recalibration, while for the device with 10% scatterer density, this ratio does go above unity for temperature shifts above 2°C. While the first two temperature variations are well within the limit of most climate-controlled rooms (even using domestic climate control systems) this highest-resolution device would likely require additional thermal control systems. However, here we note that a 1°C variation is well within the capability of low-cost thermo-electric cooling elements.

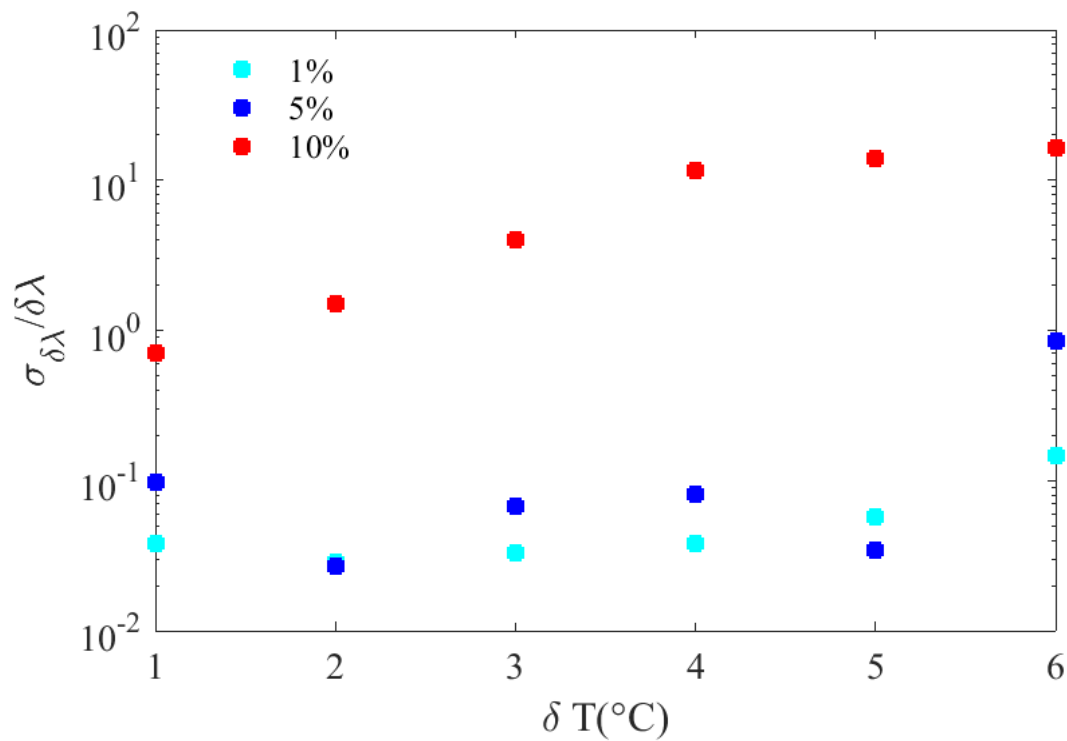

Fig. 9. Ratio of standard deviation in wavelength shift to resolving power of corresponding devices is plotted as a function of increasing temperature for devices having air holes density of 1%, 5 % and 10 %.

## 4. Conclusion

Both the throughput of an on-chip speckle spectrometer and its resolution exhibit a strong dependence on the scattering strength of the device, with a clear trade-off relationship between these two figures of merit. This relationship holds through a transition from a single scattering regime to a multiple scattering regime as the device becomes more disordered. Our work also shows that there is currently no simple stand-in figure of merit to replace the accurate characterisation of the device resolution, instead, we systematically studied the resolution of speckle spectrometers with varying disorder density and showed tuning of the device resolution from 2.1 nm to 20 pm, to our knowledge the best resolution to date in an integrated speckle spectrometer, achieved with a relevant device area below 0.0025 mm$^2$. Furthermore, we showed that a speckle spectrometer calibrated at a reference temperature has a usable temperature range, over which it can operate without the need for additional reference matrices, with this temperature range again trading off with the device resolution. For resolutions on the order of a few nm to 100's of pm this temperature range extends over several degrees within the control provided by domestic or laboratory-grade climate control systems, while for our highest resolution device, a 1°C temperature accuracy is required, achievable through commercial thermo-electric cooling elements. We can, therefore, conclude that in the case of speckle spectrometers for real-world applications, it will be essential to balance these trade-offs finely. The ideal speckle spectrometer will have just the required resolution to maximise the device throughput and operating temperature range.

## 5. Materials and Methods

### 5.1. Sample fabrication

We fabricated identical spectrometer layouts incorporating the 10 different scattering density configurations with a uniform random distribution to compare the spectrometer performance of the different scattering density devices. A MATLAB code was developed to generate 2D uniformly distributed disordered patterns with a surface density of air holes ranging from 1 % to 10 %. At 1 % scatterer density, the number of scatterers are 550, with higher densities containing corresponding multiples of this value. These scatterers were positioned on a semicircular surface with a diameter of 50 $\mu$m, using a uniform probability distribution (Mersenne Twister pseudo-random number generator). Each point in the distribution becomes the centre for an air hole with a diameter of 150 nm. The algorithm also forces a minimum edge-to-edge distance of 50 nm between each pair of holes as a fabrication constraint. Electron beam lithography is performed using a Raith 30KV e-beam system using the positive tone AR-P 6200 resist. The exposure is followed by pattern transfer into the top silicon layer of a 220 nm Si on 3 $\mu$m silica silicon-on-insulator wafer using an SF6/CHF3 reactive ion etch. A scanning electron microscope (SEM) image of the device is shown in Figure 1. As shown in the inset of Figure 1(c) the spectrometer also features a photonic-crystal mirror at the bottom to suppress reflections, an outer air trench to scatter the transmitted speckle pattern for collection using a vertically mounted objective and NIR camera, and an incoupling waveguides All these elements are defined during the same electron-beam lithography exposure as the scatterer distributions.

The photonic crystal (PhC) mirror features a complete bandgap in the wavelength region of interest (1550-1575 nm) and consists of a triangular lattice of air holes with a lattice period of $a$ = 505 nm and a hole radius of 180 nm.

### 5.2. FDTD Simulation

The FDTD simulations were conducted using Lumerical to analyze light scattering and throughput in a silicon-on-insulator (SOI) structure. A broadband Gaussian beam source was used for excitation, with perfectly matched layers (PMLs) applied at the simulation boundaries to prevent non-physical reflections. For in-plane field characterization, a field monitor was positioned at the center of the silicon-on-insulator (SOI) layer. To capture the out-of-plane scattered light, an additional monitor was placed at a vertical distance of 1 µm above the top surface of the structure. The number of scatterers per device realization was varied in discrete multiples of 90(for 1 % scatterer density). A minimum mesh step size of 0.25 nm was used to ensure accurate field resolution. For throughput analysis, the out-of-plane scattered field originating from the air trench was numerically integrated.

## 3. Characterisation

The devices were characterised using a tunable, near-infrared laser (NIR) (TS-100, EXFO), coupled to a single-mode, polarization-maintaining fiber. The fiber output is collimated using an aspheric lens, passed through a polarizer to get the pure TE component of the incoming light in the wavelength range of 1550 nm to 1565 nm, and then focused onto the cleaved edge of the access waveguide (400 nm width) using a second aspheric lens. The light scattered from the device was imaged from above using a 50X objective (Mitutoyo NIR Infinity Corrected) and an InGaAs camera (Raptor, Owl 640 II). All the experiments are performed at a stable room temperature using a Thorlabs Heater and TEC Temperature Controller (TC300B) equipped with a flexible Resistive Foil Heater (HT10K), on a floated optical table. At each scanned wavelength, the intensity distribution along the outer air trench (the region between the red arcs in Figure 3(d)) is divided into 25 equal segments. This process generates transmission matrices (TMs) with dimensions of [25 × 300] for the first seven devices, with air holes density range of 1-7 %, and [25 × 1500] for the last three devices with air holes density range 8-10%. Each column of the TM serves as a unique fingerprint corresponding to a specific wavelength, enabling the identification of unknown wavelengths. By integrating the total power within each column of the TM, the wavelength-dependent throughput of each spectrometer is determined.

## 5.5. Thermal characterisation

The temperature characterisation of the on-chip spectrometer was performed using a Thorlabs heater and TEC Temperature Controller (TC300B) equipped with a flexible Resistive Foil Heater (HT10K), ensuring temperature stability within ±0.1°C. Each device was initially characterized at room temperature by recording its characteristic transmission matrix (TM). Subsequently, the temperature was increased in 1°C increments, allowing the device to stabilize at each set temperature for 15 minutes before recording the corresponding probe matrix. Probe matrices were obtained for three devices with air hole densities of 1 %, 5 %, and 10 %, at six different temperatures, each spaced 1°C apart. The wavelength shift at each temperature was averaged over 50 measurements, referenced against the calibration matrix recorded at room temperature.

**Acknowledgments**

SAS and BK acknowledge funding from the Engineering and Physical Sciences Research Council (EPSRC) project EP/V029975/1. LDN acknowledges the support from the National Science Foundation (Grant No. ECCS-2110204).

**Disclosures**

"The authors declare no conflicts of interest."

**Data availability**

Upon publication, the underlying data will be made available through the University of St Andrews repository. NOTE FOR REVIEWERS and EDITORS: after acceptance, a doi and URL will be inserted here.